\documentclass[aps,reprint,prd,nofootinbib,noshowkeys,longbibliography,floatfix,superscriptaddress,preprintnumbers]{revtex4-1}
\usepackage[utf8]{inputenc} 
\usepackage{hyperref} 
\usepackage{amsmath} 
\usepackage{amsfonts}
\usepackage{amsthm}
\usepackage{amssymb}
\usepackage{graphicx} 
\usepackage{color}
\usepackage[sort&compress]{natbib}
\usepackage[normalem]{ulem}

\DeclareMathOperator{\tr}{Tr}

\begin{document}

\title{The Electroweak Sphaleron in a strong magnetic field}
\date{May 6, 2020}

\author{David L.-J. Ho}
\email{d.ho17@imperial.ac.uk}
\author{Arttu Rajantie}
\email{a.rajantie@imperial.ac.uk}
\affiliation{Department of Physics, Imperial College London, SW7 2AZ, UK}
\preprint{IMPERIAL-TP-2020-DH-01}

\begin{abstract}
In an external magnetic field, the energy of the electroweak sphaleron---representing the energy barrier to baryon and lepton number violation---decreases but remains nonzero until the upper Ambj{\o}rn-Olesen critical field strength set by the Higgs mass and the electric charge, where it vanishes. We demonstrate this by numerically computing the sphaleron configuration in the presence of an external magnetic field, over the full range of field strengths until the energy barrier vanishes. We discuss the implications for baryogenesis in the early universe, and the possibility of observing of baryon and lepton number violation in heavy-ion collisions.
\end{abstract}

\maketitle

\section{Introduction}
The electroweak sphaleron~\cite{manton1983topology, manton1984saddle} is a static, unstable solution to the field equations of electroweak theory; a saddle point of the energy functional. This saddle point can be interpreted as the peak energy configuration along an noncontractible loop in field configuration space connecting topologically distinct vacua. It has been shown~\cite{tHooft1976symmetry, tHooft1976computation} that due to the chiral anomaly, transitions between such vacua are associated with a violation of baryon number \(B\) and lepton number \(L\). The energy 
$E_{\rm sph}$
of the sphaleron configuration therefore represents the barrier to \(B + L\) violation within the Standard Model.
At nonzero temperature $T$, the rate of these processes is suppressed by
$\exp(-E_{\rm sph}/T).$

Previous numerical evaluations of the sphaleron solution in zero or weak external field have shown that it has a significant magnetic dipole moment~\cite{manton1984saddle, james1992sphaleron, klinkhamer1992sphaleron, kunz1992sphaleron,comelli11999sphaleron}. This suggests that the energy barrier to sphaleron transitions is lowered in the presence of a weak external magnetic field. In this paper we study this phenomenon beyond the weak-field limit by numerically computing sphaleron solutions over the full range of relevant magnetic field strengths. For strong magnetic fields nonlinear effects dominate; we use nonperturbative lattice techniques to allow us to compute sphaleron solutions in this regime.

The study of strong magnetic fields in the Standard Model is nontrivial due to the phenomenon of Ambj{\o}rn-Olesen condensation~\cite{ambjorn1988antiscreening, ambjorn1988electroweak, ambjorn1989condensate}. Ambj{\o}rn and Olesen found two critical values of the magnetic field:
\begin{align}
    \label{equ:B1value}
    B_\mathrm{crit}^{(1)} &= \frac{m_\mathrm{W}^2}{e}
    \approx 2.1 \times 10^4~{\rm GeV}^2
    \approx 1.1\times 10^{20}~{\rm T}, \\
    \label{equ:B2value}
    B_\mathrm{crit}^{(2)} &= \frac{m_\mathrm{H}^2}{e}
    \approx 5.2 \times 10^4~{\rm GeV}^2
    \approx 2.7\times 10^{20}~{\rm T},
\end{align}
where \(m_\mathrm{W}\) and \(m_\mathrm{H}\) denote the W boson and Higgs masses respectively, and \(e\) denotes electric charge. At the lower critical field, \(B_\mathrm{crit}^{(1)}\), a homogeneous magnetic field ceases to be the ground state, and is classically unstable in favour of a lattice of vortices. As the external field increases in strength above \(B_\mathrm{crit}^{(1)}\), the mean value of the Higgs field magnitude decreases continuously, until at \(B_\mathrm{crit}^{(2)}\) the symmetry of the Higgs vacuum is restored.

It has been shown~\cite{damgaard1991anomalous} that in the symmetric Higgs phase there is the potential for unsuppressed \(B + L\) violation, meaning that strong magnetic fields could provide the first observations of this elusive phenomenon. However, the nature of the sphaleron in fields approaching \(B_\mathrm{crit}^{(1)}\) and \(B_\mathrm{crit}^{(2)}\) has not previously been studied. In this paper we compute sphaleron solutions for Standard Model parameters in an external field \(B_\mathrm{ext}\) ranging from zero to \(B_\mathrm{crit}^{(2)}\). We confirm that the sphaleron energy vanishes precisely at \(B_\mathrm{ext} = B_\mathrm{crit}^{(2)}\) and not below.

\section{Theory} \label{sec:theory}
\subsection{Weinberg-Salam electroweak theory}

In this paper we study the Weinberg-Salam theory of electroweak interactions comprising part of the Standard Model of particle physics~\cite{weinberg1967model, salam1968elementary}. The sector of the Standard Model Lagrangian of interest here is the electroweak and Higgs sector, consisting of an SU(2) gauge field \(W_\mu^a\), a U(1) hypercharge gauge field \(Y_\mu\), and a scalar Higgs doublet \(\phi\):
\begin{widetext}
\begin{equation} \label{eq:electroweakLagrangian}
    \mathcal{L}_\mathrm{EW} = -\frac{1}{2} \tr (W_{\mu \nu} W^{\mu \nu}) - \frac{1}{4} Y_{\mu \nu} Y^{\mu \nu} + (D_\mu \phi)^\dagger (D^\mu \phi) - V(\phi),
\end{equation}
\end{widetext}
where
\begin{align}
    W_{\mu \nu}^a &= \partial_\mu W_\nu^a - \partial_\nu W_\mu^a + i g \varepsilon^{abc} W_\mu^b W_\nu^c, \\
    Y_{\mu \nu} &= \partial_\mu Y_\nu - \partial_\nu Y_\mu, \\
    D_\mu &= \partial_\mu + \tfrac{1}{2} i g W_\mu^a \sigma^a + \tfrac{1}{2} i g' Y_\mu, \\
    V(\phi) &= \lambda \left(\phi^\dagger \phi - v^2/2 \right)^2;
\end{align}
\(\sigma^a\) denote the Pauli matrices. There are three dimensionless parameters in the theory: the SU(2) gauge coupling \(g\), the U(1) gauge coupling \(g'\), and the Higgs self-coupling \(\lambda\). The scale is set by the Higgs vacuum expectation value (VEV) \(v / \sqrt{2}\).

After spontaneous symmetry breaking it is useful to define the weak mixing angle
\begin{equation}
    \tan \theta_\mathrm{W} = \frac{g'}{g}.
\end{equation}
The theory has three massive gauge bosons: the W bosons
\begin{equation}
    W^\pm_\mu = W^1_\mu \pm i W^2_\mu
\end{equation}
have mass \(m_\mathrm{W} = \tfrac{1}{2} g v\), and the Z boson
\begin{equation}
    Z_\mu = W_\mu^3 \cos \theta_\mathrm{W} - Y_\mu \sin \theta_\mathrm{W}
\end{equation}
has mass \(m_\mathrm{Z} = m_\mathrm{W} / \cos \theta_\mathrm{W}\). The photon
\begin{equation}
    A_\mu = W_\mu^3 \sin \theta_\mathrm{W} + Y_\mu \cos \theta_\mathrm{W}
    \label{equ:photondefcont}
\end{equation}
remains massless. The electric charge is given by
\begin{equation}
    e = g \sin \theta_\mathrm{W}\approx 0.303.
\end{equation}
Finally, the Higgs field gains a mass \(m_\mathrm{H} = \sqrt{2 \lambda} v\). The physical parameters of the Standard Model are well known~\cite{tanabashi2018review}:
\begin{align}
    m_\mathrm{W} &\approx 80.4 \ \mathrm{GeV}, \\
    m_\mathrm{H} &\approx 125.2 \ \mathrm{GeV}, \\
    \sin^2 \theta_\mathrm{W} &\approx 0.23.
\end{align}

\subsection{The electroweak sphaleron}
The electroweak sphaleron is a static, unstable solution to the classical field equations of electroweak theory~\cite{manton1983topology, manton1984saddle}. It was shown in Ref.~\cite{manton1984saddle} that the sphaleron is the maximal energy configuration along a noncontractible loop in field configuration space, and that a path traversing this loop is associated with an integer change in the Chern-Simons number. Due to the chiral anomaly~\cite{tHooft1976symmetry, tHooft1976computation}, this means that classical or quantum transitions through the sphaleron result in baryon and lepton number violation. A pedagogical introduction to the sphaleron and to electroweak baryogenesis in general can be found in Ref.~\cite{white2016pedagogical}.

In the limiting case where the weak mixing angle \(\theta_\mathrm{W}\) vanishes, the sphaleron is spherically symmetric. However, at finite values of \(\theta_\mathrm{W}\) the sphaleron solution has a dipole moment~\cite{manton1984saddle, james1992sphaleron, klinkhamer1992sphaleron, kunz1992sphaleron}. According to Ref.~\cite{hindmarsh1993origin} the main contribution to this dipole moment can be interpreted as coming from a small segment of Z string~\cite{nambu1977string, vachaspati1992vortex}, which terminates on a Nambu monopole-antimonopole pair. The topological nature of the sphaleron may be interpreted as a relative ``twist'' between the monopole and antimonopole~\cite{vachaspati1994electroweak, Achucarro1999semilocal}. There is also a small contribution to the sphaleron dipole moment from a loop of electromagnetic current density~\cite{hindmarsh1993origin}.

The fact that the electroweak sphaleron has a nonzero dipole moment for the physical value of the weak mixing angle indicates that its energy may be lowered by the presence of a suitably aligned external magnetic field. For weak external fields, one can assume that this effect is linear: the change in energy is given by
\begin{equation} \label{eq:linearSphaleronEnergy}
    \Delta E_\mathrm{sph} = -\vec{B}_\mathrm{ext} \cdot \vec{\mu}_\mathrm{sph},
\end{equation}
where \(\vec{B}_\mathrm{ext}\) denotes the external magnetic field and \(\vec{\mu}_\mathrm{sph}\) denotes the sphaleron dipole moment. However, for stronger magnetic fields nonlinear effects become important; these cannot be calculated analytically. A numerical study of the sphaleron in an external magnetic field was carried out by Comelli \emph{et al.} in the weak-field regime~\cite{comelli11999sphaleron}. However, this analysis only considered solutions where the fields did not have any angular dependence, so could not be extended far into the nonlinear regime. In this work we have carried out a numerical analysis over the full range of physically interesting external magnetic fields.

\subsection{Ambj{\o}rn-Olesen Condensation}

The study of the sphaleron in a strong magnetic field is made both more interesting and more complicated by the fact that for sufficiently high field strengths, the ground state of electroweak theory becomes nontrivial. This effect was first studied in detail by Ambj{\o}rn and Olesen~\cite{ambjorn1988electroweak, ambjorn1988antiscreening, ambjorn1989condensate}. Due to the interaction between the charged vector bosons and the external field, at high enough external field strengths it becomes energetically favourable for a vector boson condensate to form, leading to the formation of a periodic lattice of vortices. The homogeneous vacuum becomes unstable at a field strength $B_\mathrm{crit}^{(1)} = m_\mathrm{W}^2/e$ set by the W boson mass.
As the magnetic field strength increases beyond this first critical field, the deviation from the homogeneous vacuum increases in magnitude, and the mean magnitude of the Higgs field decreases. This continues until the external field strength reaches a second critical value
$    B_\mathrm{crit}^{(2)} = m_\mathrm{H}^2/e.$
For field strengths above this it is energetically favourable for the Higgs field to be in the symmetric phase, and the magnetic field is pure hypercharge.
This symmetry restoration has important consequences for the electroweak sphaleron. 

In the symmetric phase of the Higgs field the sphaleron field configuration is pure gauge, so it is expected that the energy of the sphaleron will be zero at this point. However, it is not immediately obvious that the sphaleron energy remains nonzero all the way up to $B_\mathrm{crit}^{(2)}$. In recent work~\cite{ho2020classical} we showed that in a theory admitting 't~Hooft--Polyakov magnetic monopoles, there exists a classical mode of monopole-antimonopole pair production at \(B_\mathrm{ext} = B_\mathrm{crit}^{(1)}\). The calculation in Ref.~\cite{ho2020classical} also involved finding a saddle point solution of classical field equations, containing a monopole-antimonopole pair and a current ring. It is therefore reasonable to ask if sphaleron processes become unsuppressed at this energy in the electroweak theory, too---perhaps due to Ambj{\o}rn-Olesen vortices carrying baryon number.

Another possibility for vanishing sphaleron energy below \(B_\mathrm{crit}^{(2)}\) is a regime where Z strings are (dynamically) stable. This was analysed in Ref.~\cite{garriga1995stability} for the unphysical case where \(m_\mathrm{H} < m_\mathrm{Z}\). It was found that an infinite Z string is perturbatively stable above some field strength smaller than \(B_\mathrm{crit}^{(1)}\). As it has been shown that Z strings can carry baryon number~\cite{vachaspati1992vortex} (though presumably not without an energy cost), an analogous phenomenon in a system with physical parameters could also provide a \(B+L\) violation mechanism.

\subsection{Lattice discretisation} \label{sec:discretisation}
As the sphaleron is a static solution, we can restrict the problem to three spatial dimensions, and set the timelike components of the gauge fields to zero. In order to perform numerical calculations we discretise the electroweak Lagrangian \eqref{eq:electroweakLagrangian}, defining a 3D lattice of points \(\vec{x} = (n_x, n_y, n_z)a\) where \(n_x, n_y, n_z\) are integers and \(a\) is the lattice spacing. The Higgs field \(\phi(\vec{x})\), which is a complex doublet, is defined on lattice points, whilst the gauge fields are defined by link variables \(U^\mathrm{W}_i(\vec{x})\in {\rm SU}(2)\) and  \(U^\mathrm{Y}_i(\vec{x})\in {\rm U}(1)\). The discretised energy density is then
\begin{widetext}
\begin{equation} \label{eq:latticeEnergyDensity}
\begin{split}
    \mathcal{E}_\mathrm{lat} &= \frac{2}{g^2 a^4} \sum_{i < j} \left[2 - \tr U_{ij}^\mathrm{W} (\vec{x})\right] + \frac{2}{g^2 a^4 \tan^2 \theta_\mathrm{W}} \sum_{i < j} \left[1 - \mathop{\mathrm{Re}}  U_{ij}^\mathrm{Y}(\vec{x})\right] \\
    &+ \frac{2}{a^2} \sum_i \left\{\phi^\dagger(\vec{x}) \phi(\vec{x}) - \mathop{\mathrm{Re}} \left[ \phi^\dagger(\vec{x}) U_i^\mathrm{Y}(\vec{x}) U_i^\mathrm{W}(\vec{x}) \phi(\vec{x} + \hat{\imath}) \right]\right\} \\
    &+ V(\phi).
\end{split}
\end{equation}
\end{widetext}
Here \(U_{ij}^\mathrm{W}\) and \(U_{ij}^\mathrm{Y}\) respectively denote the SU(2) and U(1) Wilson plaquettes. The sum of this over all lattice sites \(E_\mathrm{lat} = \sum_{\vec{x}} a^3 \mathcal{E}_\mathrm{lat}\) is the quantity we extremise.

In practice, we work in the unitary gauge,  \(\phi(x) = \begin{pmatrix} 0 & h(x) \end{pmatrix}^T\), $h(x)\in\mathbb{R}$. 
Near the vacuum state the residual electromagnetic field then corresponds to the complex phase of the top left element $u_i(\vec{x})\equiv\left[U_i(\vec{x})\right]_{11}$ of the combined link variable
$U_i(\vec{x})=U_i^\mathrm{Y}(\vec{x}) U_i^\mathrm{W}(\vec{x})$. This agrees with Eq.~(\ref{equ:photondefcont}) in the continuum limit $a\rightarrow 0$. 
The electromagnetic field strength tensor is then given by the plaquette variable
\begin{equation}
    F_{ij}(\vec{x})=\frac{1}{ea^2}
    \arg u_i(\vec{x}) u_j(\vec{x} + \hat{\imath}) u_i^*(\vec{x} + \hat{\jmath}) u_j^*(\vec{x}),
\end{equation}
and the magnetic field strength in the standard way as
$B_i=\epsilon_{ijk} F_{jk}/2$.

\begin{figure}
    \centering
    \includegraphics[width=0.5\textwidth]{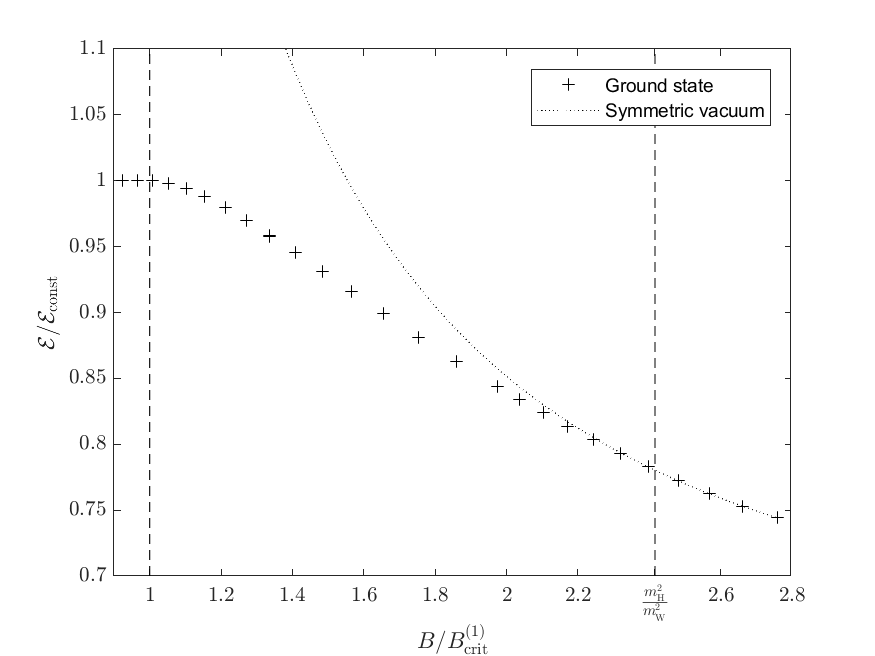}
    \caption{Plot of spatially averaged ground state energy density (divided by that of a homogeneous magnetic field \(\mathcal{E}_\mathrm{hom}\)) against external field strength in units of the lower critical field, for Standard Model parameters. The dotted curve shows energy density of a pure hypercharge field in the symmetric Higgs phase. Dashed lines indicate the lower and upper critical fields.}
    \label{fig:vacuumEnergyDensity}
\end{figure}

A useful test of the lattice discretisation is computation of the ground state energy density in the Ambj{\o}rn-Olesen phase. This was carried out using gradient flow on a \(64 \times 64\) grid (the solution is translation-invariant along the axis parallel to the external field). An external magnetic field was enforced using suitable boundary conditions (more detail can be found in Section~\ref{sec:results}) and varied incrementally by changing the Higgs VEV. The results are shown for Standard Model parameters in Fig.~\ref{fig:vacuumEnergyDensity}, where the normalised ground state energy density is plotted against magnetic field strength. In line with the results of Refs.~\cite{ambjorn1988antiscreening, ambjorn1988electroweak, ambjorn1989condensate}, the energy density interpolates between that of a homogeneous magnetic field in the broken Higgs phase for \(B_\mathrm{ext} / B_\mathrm{crit}^{(1)} < 1\), and a pure hypercharge field in the symmetric Higgs phase for \(B_\mathrm{ext} / B_\mathrm{crit}^{(1)} > m_\mathrm{H}^2 / m_\mathrm{W}^2 \approx 2.42\).

\section{Electroweak sphaleron in an external magnetic field} \label{sec:results}

\begin{figure}
    \centering
    \includegraphics[width=0.5\textwidth]{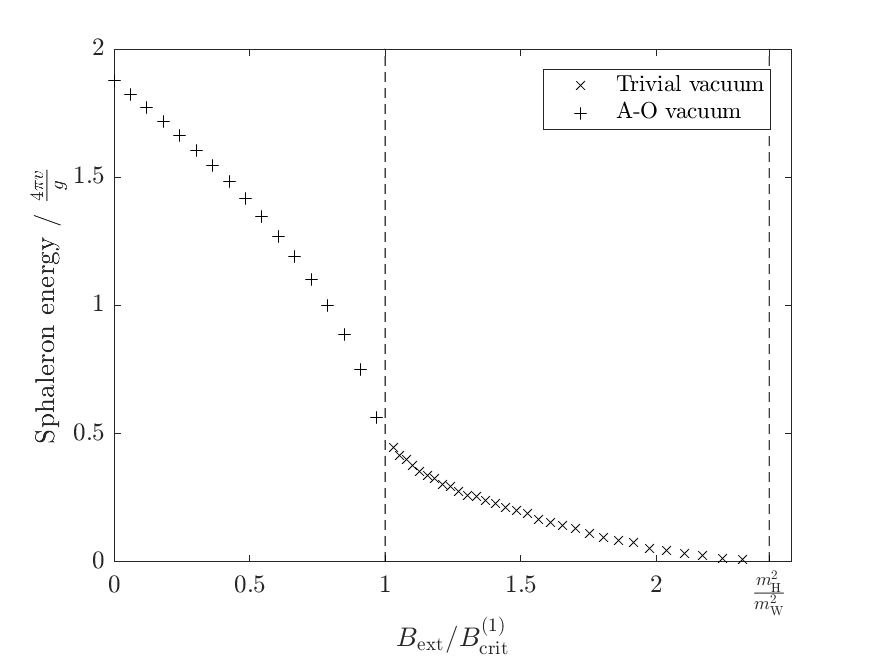}
    \caption{Plot of sphaleron energy (with the background field subtracted) against external field strength for physical values of the Standard Model parameters. Dashed lines indicate the lower and upper critical fields. In physical units, \(4 \pi v / g \approx 5 \ \mathrm{TeV}\).}
    \label{fig:sphaleronEnergyPlot}
\end{figure}

The numerical methods used in this paper are similar to those used in Ref.~\cite{ho2020classical}: We employed a modified gradient flow first proposed in Ref.~\cite{chigusa2019bounce} to converge on saddle points of the discretised energy. For full numerical details, readers are invited to consult these papers, but we include a brief summary of the method in Appendix~\ref{appendix:modifiedGradientMethod}. The method we employ has also recently been used in Ref.~\cite{hamada2020sphaleron} to compute the sphaleron in the limiting case \(\theta_\mathrm{W} = 0\), without an external magnetic field present. The \textsc{LAT}field2 C++ library~\cite{daverio2015latfield} was used to aid parallelisation and a Barzilai-Borwein adaptive step size~\cite{barzilai1988step} was used to speed convergence.

To find the sphaleron solution without an external field, we used the initial ansatz described in Ref.~\cite{klinkhamer1992sphaleron}. As detailed in Ref.~\cite{ho2020classical}, periodic boundary conditions constrain the magnetic flux through the lattice in units of \(4 \pi / e\); this can be exploited to restrict the gradient flow to field configurations with an external magnetic field present. By linearly superposing a constant field on an existing sphaleron solution and using this as an initial condition for modified gradient flow, sphalerons in an external field were iteratively generated until the field strength was within one flux quantum of the first critical field \(B_\mathrm{crit}^{(1)}\).

For \(B_\mathrm{ext} > B_\mathrm{crit}^{(1)}\), a sphaleron solution with a nontrivial background field is sought. This was achieved by first using standard gradient flow to find the Ambj{\o}rn-Olesen vortex background, then ``transplanting'' a sphaleron solution with an incrementally weaker external field by replacing the field at lattice points outside the sphaleron core with the Ambj{\o}rn-Olesen vortex field. After smoothing using standard gradient flow, the modified gradient flow of Ref.~\cite{chigusa2019bounce} could be used to find a saddle-point solution of the fields that tends to the Ambj{\o}rn-Olesen vortex lattice at large distances. After the first sphaleron solution against an Ambj{\o}rn-Olesen background was found, the magnitude of the external field was increased further by changing the scalar field VEV. This was continued until the external field reached \(B_\mathrm{ext} = B_\mathrm{crit}^{(2)}\).

Computation of the sphaleron for \(B_\mathrm{ext} < B_\mathrm{crit}^{(1)}\) was carried out using a \(64 \times 64 \times 192\) grid; for \(B_\mathrm{ext} > B_\mathrm{crit}^{(1)}\) a \(64 \times 64 \times 256\) grid was used. Table~\ref{tab:parameterTable} gives the values or ranges of the dimensionless parameters of Eq.~\eqref{eq:latticeEnergyDensity} used in our calculations. Note that the overall scale is set by the lattice spacing \(a\), and the boson mass hierarchy is set by the ratios \(\lambda / g^2\) and \(\sin^2 \theta_\mathrm{W}\), so matching these quantities to the Standard Model parameters ensures that our results reflect the physical Standard Model.

\begin{table}
    \centering
    \begin{tabular}{c c}
        \hline \hline
        Parameter & Value \\ \hline 
        \(g\) & 0.5 \\
        \(\lambda\) & 0.076 \\
        \(a v\) & 0.60-0.90 \\
        \(\sin^2 \theta_\mathrm{W}\) & 0.23 \\
        \hline \hline
        
    \end{tabular}
    \caption{Numerical values of the dimensionless parameters of Eq.~\eqref{eq:latticeEnergyDensity} used in our calculations.}
    \label{tab:parameterTable}
\end{table}

\subsection{Weak magnetic fields}

\begin{figure*}
    \centering
    \begin{tabular}{c@{}c@{}c@{}}
        \includegraphics[width=0.33\textwidth]{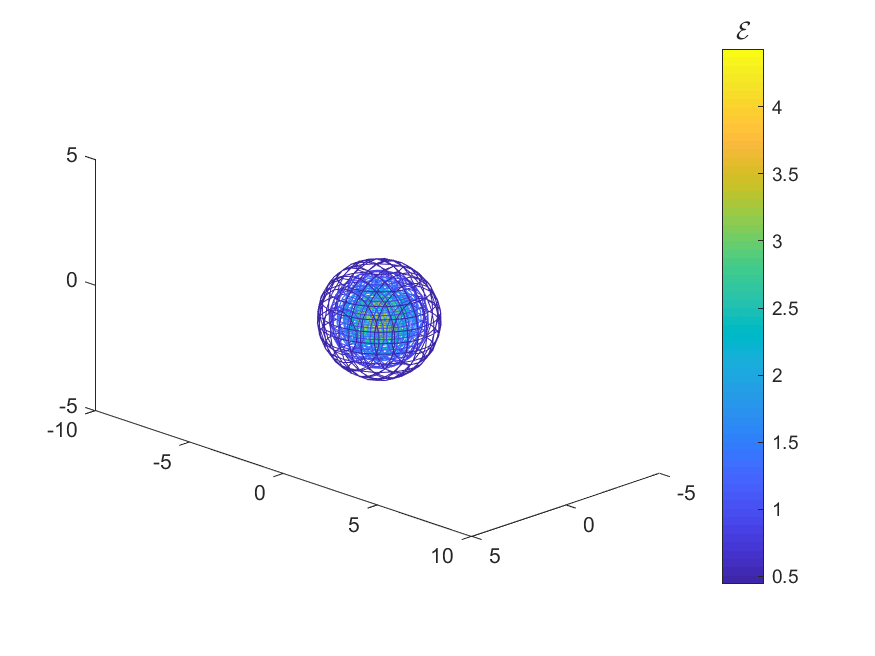} &
        \includegraphics[width=0.33\textwidth]{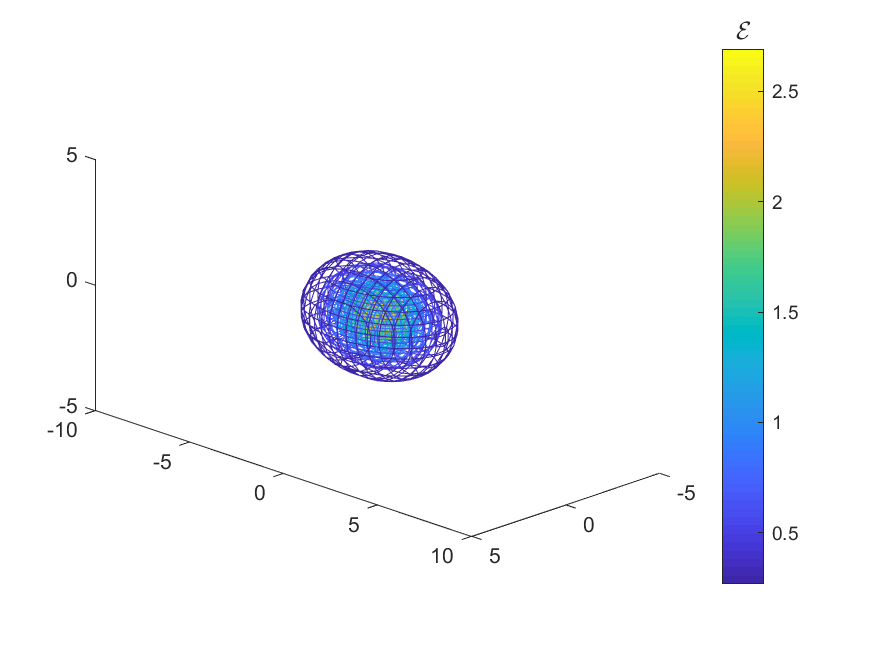} &
        \includegraphics[width=0.33\textwidth]{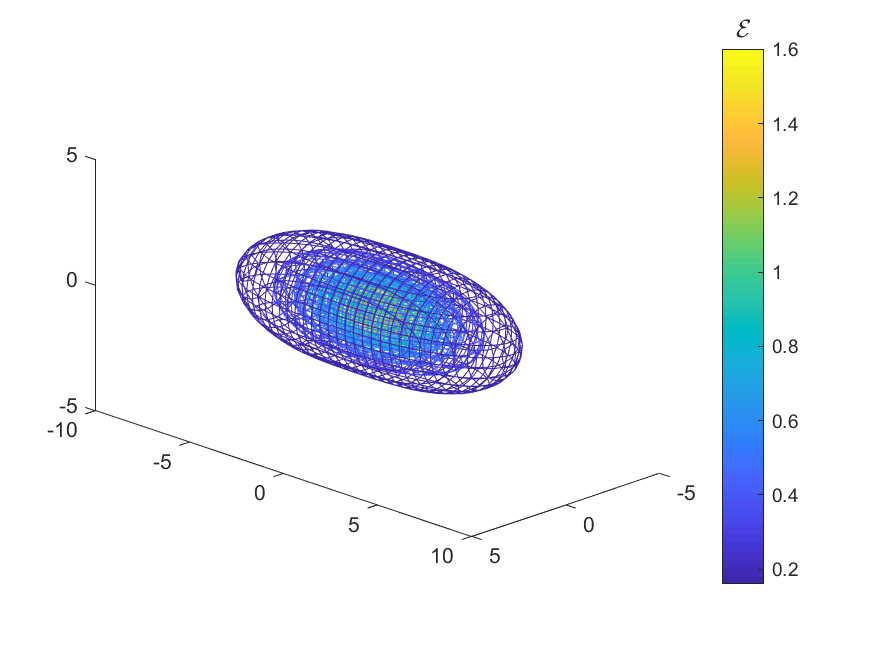} \\
        \includegraphics[width=0.33\textwidth]{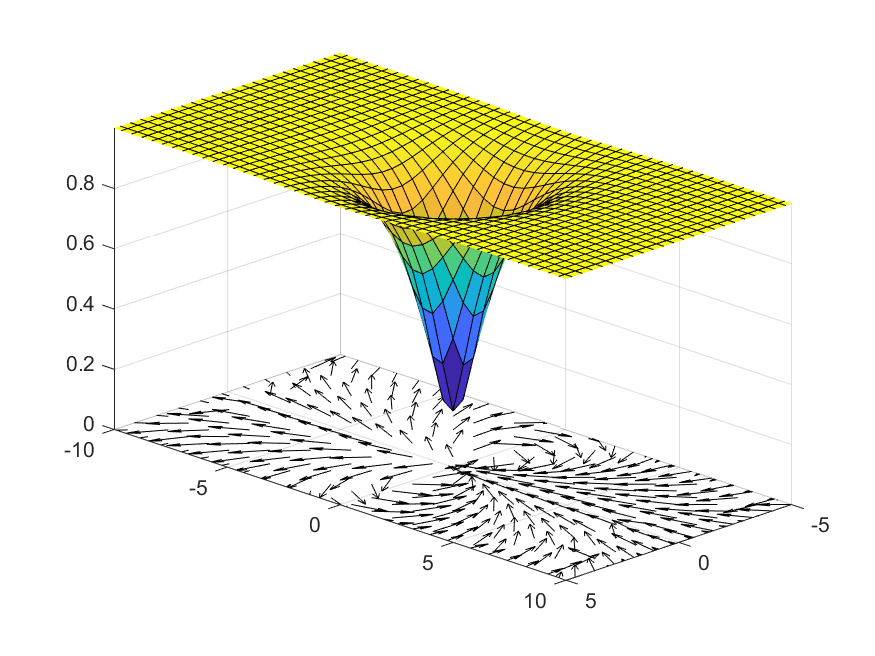} &
        \includegraphics[width=0.33\textwidth]{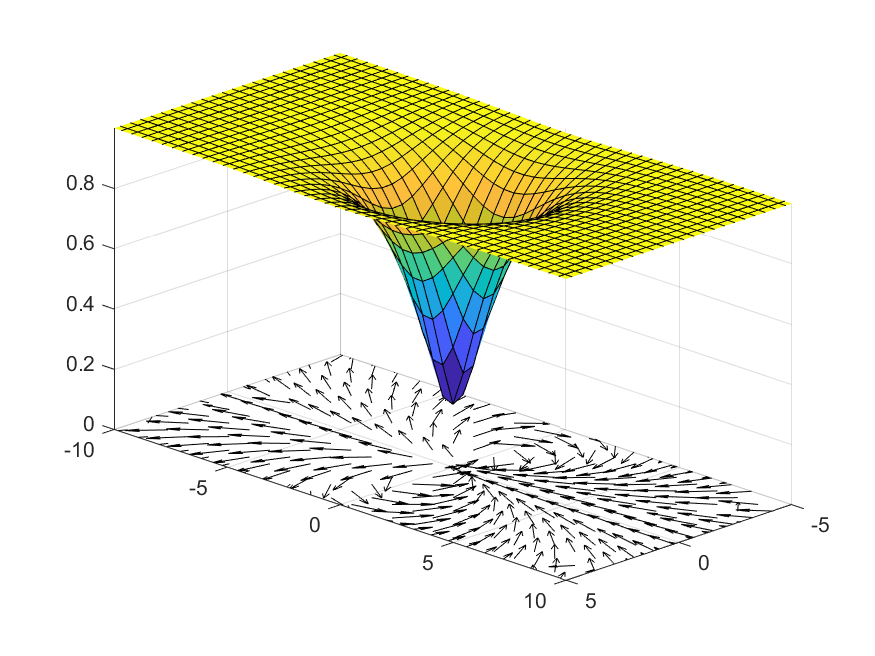} &
        \includegraphics[width=0.33\textwidth]{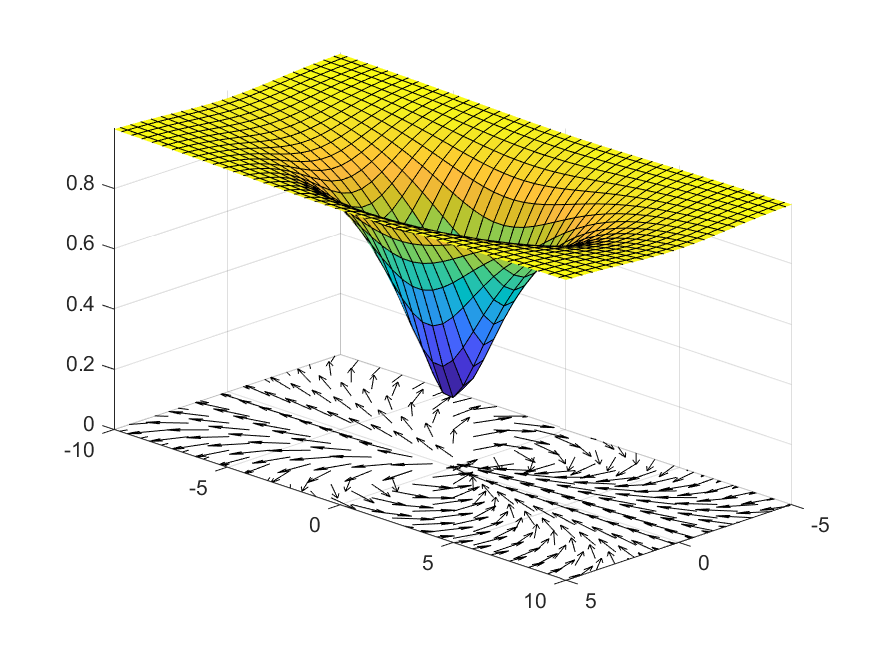} \\
        {(}a{)} \(B/B_\mathrm{crit}^{(1)} = 0\) &
        {(}b{) \(B/B_\mathrm{crit}^{(1)} = 0.48\)} &
        {(}c{) \(B/B_\mathrm{crit}^{(1)} = 0.91\)}
    \end{tabular}
    \caption{Visualisations of sphaleron solutions for different subcritical magnetic field strengths. Top plots show energy density contours in units of \((gv)^4\) in 3D space. Bottom plots show slices in a plane parallel to the magnetic field intersecting the sphaleron at its centre: the surface gives the scalar field magnitude in units of its VEV, whilst the vector plots give the direction of the magnetic field (with the background subtracted) through the same slice. Spatial axes have units of \((gv)^{-1}\).}
    \label{fig:sphaleronPlots}
\end{figure*}

The sphaleron energy as a function of \(B_\mathrm{ext}\) is shown in Fig.~\ref{fig:sphaleronEnergyPlot}. For weak fields \(B_\mathrm{ext} \ll  B_\mathrm{crit}^{(1)}\), there is a linear relationship as expected from Eq.~\eqref{eq:linearSphaleronEnergy}. The gradient in the linear region suggests a dipole moment of \(\mu \approx 1.8 \, e/(\alpha_\mathrm{W} m_\mathrm{W})\) (\(\alpha_\mathrm{W}\) denoting the weak fine structure constant), which agrees with interpolated results from other numerical studies of the sphaleron dipole moment~\cite{kunz1992sphaleron, klinkhamer1992sphaleron, comelli11999sphaleron} (these were carried out before the Higgs mass was known, so do not include data for the physical Higgs mass).

As the external field strength approaches \(B_\mathrm{crit}^{(1)}\) from below, the sphaleron energy decreases at an increasing rate. Interpolating the values in Fig.~\ref{fig:sphaleronEnergyPlot} gives a sphaleron energy at the critical field of \(E_\mathrm{sph} (B_\mathrm{crit}^{(1)}) \approx 2 \pi v / g \approx 2.5 \ \mathrm{TeV}\), significantly smaller than than the sphaleron energy in the absence of an external field but significantly above zero.

Visualisations of sphaleron solutions for \(B_\mathrm{ext} < B_\mathrm{crit}^{(1)}\) are shown in Fig.~\ref{fig:sphaleronPlots}. It can be clearly seen that the solution becomes more prolate as the field increases: for an external field of around half the first critical field the contours are still close to spherical (as the weak mixing angle is small, the prolation of the sphaleron in the absence of an external field is barely noticeable). However, the prolation becomes very pronounced at larger field strengths. It can be seen from Fig.~\ref{fig:sphaleronPlots} that the peak energy density of the sphaleron decreases monotonically with increasing field strength. The contours of the Higgs field magnitude also become more prolate, though there always remains a minimum close to zero at the centre of the sphaleron---this is due to the topologically nontrivial nature of the solution; in the continuum limit the Higgs field would vanish at the centre. The lower plots in Fig.~\ref{fig:sphaleronPlots} show the direction of the magnetic field as defined in Section~\ref{sec:discretisation}. The main observable feature is the dipole field due to a pair of Nambu monopoles as observed in Ref.~\cite{hindmarsh1993origin}; the length of the segment of Z string is two lattice spacings for all \(B_\mathrm{ext} < B_\mathrm{crit}\). The magnetic part of the Z field also shows a dipole-like configuration, though this decays much more rapidly with spatial distance as the Z boson is massive. Examination of the hypercharge field shows, again as described in Ref.~\cite{hindmarsh1993origin}, a loop of (electric) current density circling the centre of the sphaleron.

\subsection{Strong magnetic fields}

\begin{figure}
    \centering
    \includegraphics[width=0.5\textwidth]{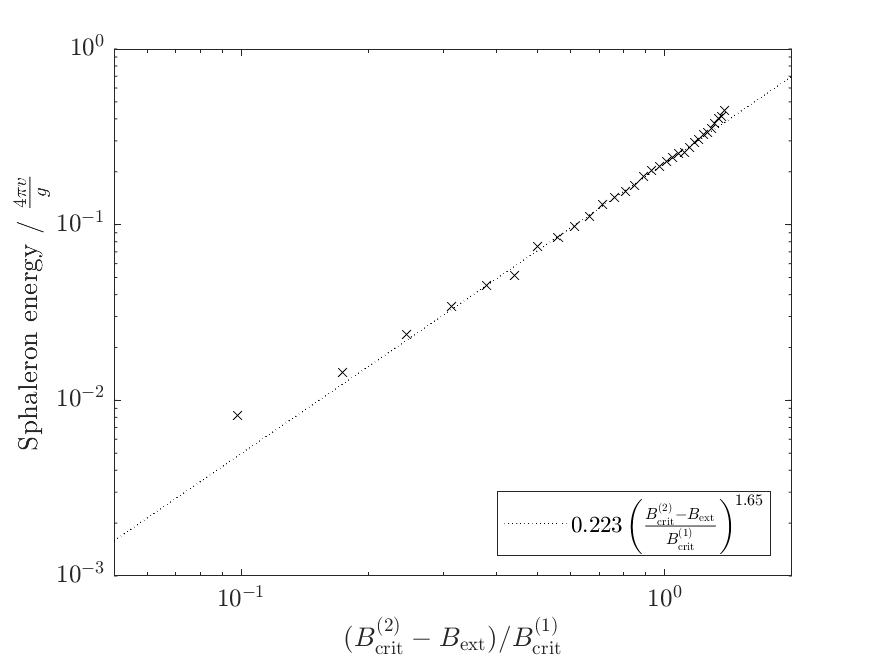}
    \caption{Logarithmic-scale plot showing sphaleron energy as the external field approaches \(B_\mathrm{crit}^{(2)}\). The dotted line shows a power law fit with the largest seven and smallest two points excluded.}
    \label{fig:strongFieldEnergyPlot}
\end{figure}

For \(B_\mathrm{ext} > B_\mathrm{crit}^{(1)}\) the homogeneous vacuum develops a negative mode and is no longer the ground state. This makes computation of the sphaleron much more difficult.

By symmetry arguments, there should be a stationary point of the energy resembling the sphaleron in subcritical fields: an axisymmetric field configuration with a Chern-Simons number of one half. An argument for the existence of such a solution follows from the argument presented in Refs.~\cite{manton1983topology, manton1984saddle}: the homogeneous vacuum is the lowest-energy axisymmetric state for a given value of \(B_\mathrm{ext}\), and one can construct a noncontractible loop from this vacuum to itself passing through a stationary point of the energy. However, such a stationary point would have two negative modes: in addition to the mode varying Chern-Simons number there would also be the instability of the background field configuration. This would mean that this solution has multiple negative modes and would not be relevant for tunnelling processes, so we have not attempted to study it here.

Instead, the relevant sphaleron solution with a single negative mode should
resemble the subcritical sphaleron at short distances but tend to the Ambj{\o}rn-Olesen vacuum at large distances. In this work we searched for and found such solutions. The key features are similar both above and below the first critical field: the energy contours trace a prolate spheroid, the Higgs field vanishes at the centre, and a pair of Nambu monopoles separated by a Z string are observed.

\begin{figure*}
    \centering
    \begin{tabular}{c@{}c@{}c@{}}
        \includegraphics[width=0.5\textwidth]{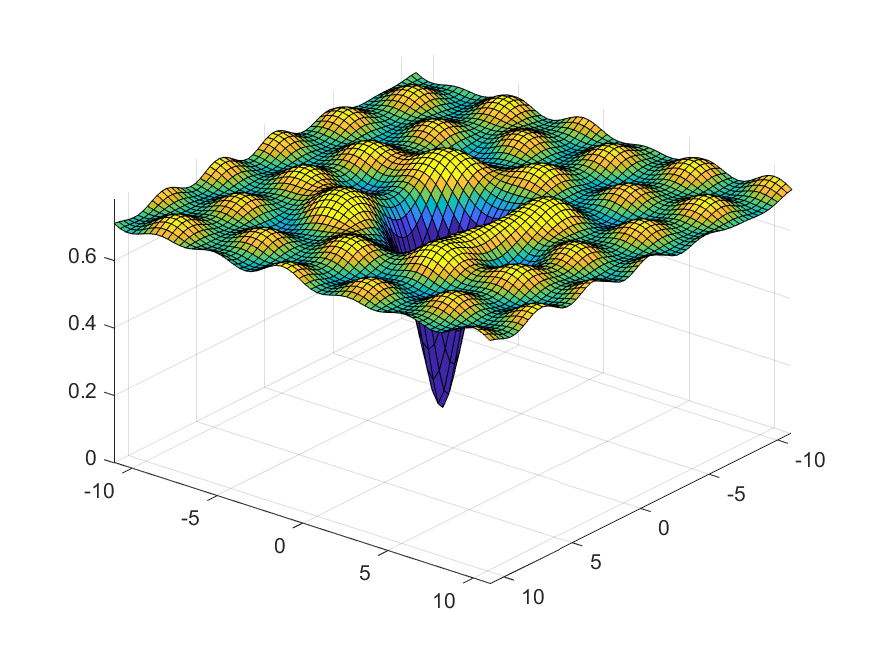} &
        \includegraphics[width=0.5\textwidth]{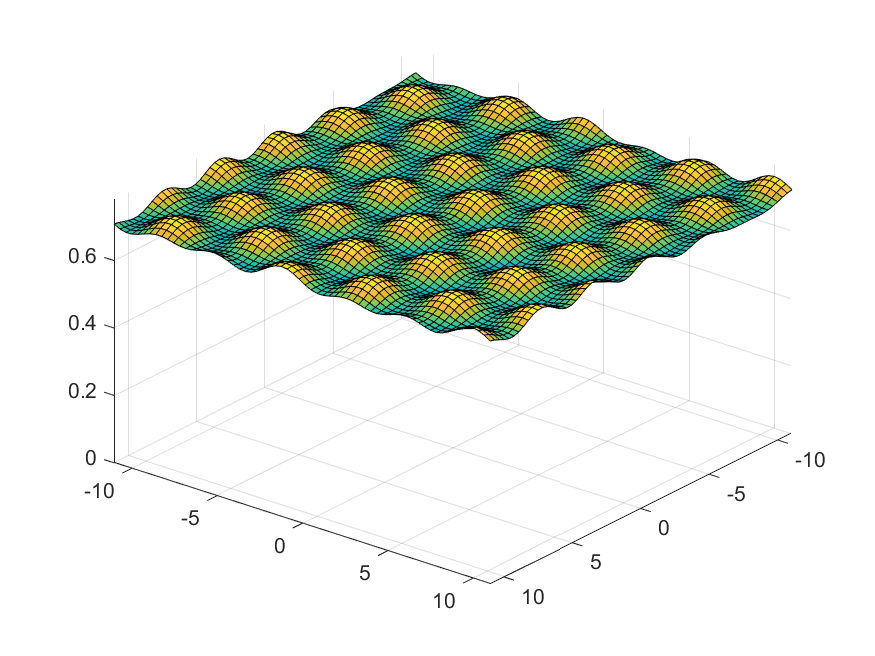} & \\
        {(}a{)} &
        {(}b{)} &
    \end{tabular}
    \caption{Surface plots of Higgs field magnitude in units of the Higgs VEV. Left: sphaleron solution for \(B_\mathrm{ext} / B_\mathrm{crit}^{(1)} = 1.70\); a slice perpendicular to the external field passing through the centre of the sphaleron is displayed. Right: Ambj{\o}rn-Olesen condensate solution for the same external field. Spatial axes have units of \((gv)^{-1}\).}
    \label{fig:AOSphaleronHiggsPlot}
\end{figure*}

A surface plot of the Higgs field magnitude for a sphaleron solution with \(B_\mathrm{ext} > B_\mathrm{crit}^{(1)}\) is shown in Fig.~\ref{fig:AOSphaleronHiggsPlot}. The Higgs field in the ground state is also shown for comparison; it is clear that the solution consists of the sphaleron of Refs.~\cite{manton1984saddle, manton1983topology} against the nontrivial background of Refs.~\cite{ambjorn1988antiscreening, ambjorn1988electroweak, ambjorn1989condensate}.

While finding solutions for external fields very close to \(B_\mathrm{crit}^{(1)}\) is technically difficult due to the presence of the almost-zero Ambj{\o}rn-Olesen mode, the points in Fig.~\ref{fig:sphaleronEnergyPlot} seem to indicate that the energy is continuous across the critical field. It can be seen that the energy of the sphaleron above the first critical field continues to decrease monotonically, though the acceleration of the decrease in energy with increasing field strength ceases. 

The sphaleron energy remains greater than zero until the second critical field \(B_\mathrm{crit}^{(2)} = m_\mathrm{H}^2 / e\) is reached. At this point, the negative mode of the sphaleron ``flattens'' to a zero mode, and standard gradient flow from a sphaleron configuration at lower external field will converge upon a field configuration with the same energy as the vacuum, with a vanishing Higgs field magnitude. This solution still contains a Z string, but with no Higgs VEV this is purely a gauge object.

To quantify the critical scaling behaviour of the sphaleron energy as the field approaches $B_\mathrm{crit}^{(2)}$, we show it on a logarithmic scale in Fig.~\ref{fig:strongFieldEnergyPlot}.
For the strongest fields the sphaleron solution is very close to the vacuum: the contribution to the total energy of the system due to the sphaleron is around one part in \(10^5\). For this reason, finding the exact solution is difficult and the value of the sphaleron energy may carry some error. This could explain the slight modulation of the curve visible on a logarithmic scale in Fig.~\ref{fig:strongFieldEnergyPlot}, especially in the smallest values of the energy. Unfortunately, as the errors are not statistical in nature, it is not possible to include error bars in this plot.

There appears to be a linear region on the log-log plot that suggests a power law scaling: for \(B_\mathrm{ext}\) close to \(B_\mathrm{crit}^{(2)}\),

\begin{equation}
\frac{g E_\mathrm{sph}}{4 \pi v} \approx 0.223 \left(\frac{B_\mathrm{crit}^{(2)} - B_\mathrm{ext}}{B_\mathrm{crit}^{(1)}} \right)^{1.65}.
\end{equation}
A numerical fit gives an exponent of \(1.65 \pm 0.04\) and a coefficient of \(0.223 \pm 0.002\). However, the errors in the data, especially for higher values of the field where the energy is very small, make the scaling relationship difficult to determine precisely. More investigation would be required to ascertain whether the power-law scaling is valid.

As the modified gradient descent method used in our calculations tends to converge to saddle points near to the initial field configuration, in principle there could be other saddle points for external fields between \(B_\mathrm{crit}^{(1)}\) and \(B_\mathrm{crit}^{(2)}\) that have a lower energy than the solutions we have found. However, our solution appears to be continuously connected to the standard zero-field sphaleron, suggesting that it is responsible for the same \(B+L\) violating process. As even the zero-field sphaleron has not been shown rigorously to be the lowest energy saddle point of electroweak theory, we do not attempt to prove this here.

\section{Conclusions}
We have computed the electroweak sphaleron solution in an external magnetic field \(B_\mathrm{ext}\) ranging from zero up to the second Ambj{\o}rn-Olesen critical field \(B_\mathrm{crit}^{(2)} = m_\mathrm{H}^2 / e\) where the Higgs field symmetry is restored. 
We find that when $B_{\rm ext}$ is increased, the sphaleron energy initially decreases linearly for 
\(B_\mathrm{ext} \ll B_{\rm crit}^{(1)} \),
then more rapidly until 
\(B_\mathrm{ext} = B_{\rm crit}^{(1)} \),
For 
$B_{\rm crit}^{(1)}< B_\mathrm{ext} < B_{\rm crit}^{(2)},$
the sphaleron energy decreases gradually until it reaches zero when the Higgs symmetry is restored
at $B_{\rm crit}^{(2)}$; above this the sphaleron ceases to exist.

The vanishing of the sphaleron energy at the second critical field confirms the result of Ref.~\cite{damgaard1991anomalous} that if such fields could be produced they would result in a greatly enhanced rate of \(B + L\) violation. A striking consequence of this is that any baryonic matter placed in such a magnetic field would spontaneously decay into leptons. 

One potential avenue for observation of this field-induced baryon and lepton number violation is heavy-ion collisions~\cite{manton2019inevitability}; these produce the strongest fields in the present-day universe~\cite{deng2012event}. As Eqs.~(\ref{equ:B1value}) and (\ref{equ:B2value}) show, the scale of the magnetic field at which the enhancement becomes significant is of order $10^4~{\rm GeV}^2$.
Numerical simulations~\cite{deng2012event, gould2019towards} indicate that the fields in present-day heavy-ion collisions at the LHC are of the order \(1 \ \mathrm{GeV^2}\), and scale linearly with energy, so collisions energies of around $10^5~{\rm TeV}$ per nucleon would be needed. This regime is inaccessible to particle colliders in the near future. In a \(10 \ \mathrm{TeV}\) collision, the reduction in sphaleron energy due to the magnetic field is only of order \(0.1 \%\), and therefore baryon number violation is almost as strongly suppressed as at zero field. In addition, at high energies the magnetic field is highly localised in both space and time~\cite{deng2012event, gould2019towards}: while according to Ref.~\cite{damgaard1991anomalous} the spatial localisation has a suppressing effect, we showed in Ref.~\cite{gould2019towards} that time dependence enhances a similar nonperturbative tunnelling phenomenon.

The results may also have relevance for cosmology.
It is possible that strong magnetic fields were present 
in the early Universe~\cite{Turner:1987bw,Joyce:1997uy,Durrer:2013pga}. If the field strength was still above or close to $B_{\rm crit}^{(1)}$ after the electroweak phase transition, it would have allowed baryon number violating sphaleron processes to continue for longer, thereby affecting the baryon asymmetry of the Universe. The required fields could be produced by exotic physics such as superconducting cosmic strings~\cite{Witten:1984eb,damgaard1991anomalous} or near-extremal magnetically charged black holes~\cite{maldacena2020comments}, but are not expected to arise in the simplest cosmological scenarios. An empirical observation, direct or indirect, of baryon number violation due to strong magnetic fields is therefore unlikely in the near future.

\section*{Acknowledgements}
The authors would like to thank Prof. Nick Manton and Prof. Jan Ambj{\o}rn for helpful discussions and comments on the manuscript.

The authors also wish to acknowledge the Imperial College Research Computing Service for computational resources.

D.L.-J.H. was supported by a U.K. Science and Technology Facilities Council studentship. A.R. was supported by the U.K. Science and Technology Facilities Council grant ST/P000762/1 and Institute for Particle Physics Phenomenology Associateship.

\appendix
\section{Modified gradient descent method} \label{appendix:modifiedGradientMethod}

Here we give a brief overview of the modified gradient descent method used in our calculations to find saddle point solutions in lattice electroweak theory. For full details, the reader is invited to consult the original paper by Chigusa \emph{et al.} introducing the method~\cite{chigusa2019bounce}.

The aim of the algorithm is to find a saddle point of the lattice energy \(E_\mathrm{lat} = \sum_{\vec{x}} a^3 \mathcal{E}_\mathrm{lat}\), where the energy density \(\mathcal{E}_\mathrm{lat}\) is defined in Eq.~\eqref{eq:latticeEnergyDensity}. If we denote the set of all field and link variables by \(X\), and an individual field or link variable by \(X_\alpha\), a standard gradient flow update may be written

\begin{equation}
    X_\alpha(\tau + \delta \tau) = X_\alpha(\tau) - \partial_\alpha E_{\rm lat} \delta \tau,
\end{equation}
where \(\tau\) denotes `flow time' and
\begin{equation} \label{eq:naiveGradientFlow}
    \partial_\alpha E_{\rm lat} := \frac{\partial E_{\rm lat}}{\partial X_\alpha}.
\end{equation}

This descends along all negative modes, so converges on minima of the energy; if the initial condition is in the vicinity of a saddle point the algorithm will move away from the saddle point with increasing flow time.

Instead of this, Chigusa \emph{et al.}'s algorithm~\cite{chigusa2019bounce} uses the update
\begin{equation} \label{eq:chigusaFlow}
    X_\alpha(\tau + \delta \tau) = X_\alpha(\tau) - \left[\partial_\alpha E_\mathrm{lat}
    - k G_\alpha \sum_\beta (\partial_\beta E_{\rm lat})G_\beta \right] \delta \tau,
\end{equation}
where \(k > 1\) is a real constant, and \(G\) is a field on the lattice, normalised such that \(\sum_\beta G_\beta G_\beta = 1\). In Ref.~\cite{chigusa2019bounce}, it is shown that any fixed point of the flow \eqref{eq:chigusaFlow} is also a stationary point of \(E_\mathrm{lat}\), and that for suitably chosen \(k\) and \(G\), this stationary point will be a saddle point.

The ideal choice for \(G\) would be directly proportional to the negative mode of the saddle point field configuration---then the modified gradient flow would descend along all positive modes and ascend along the negative mode to the sphaleron. As this is not known \emph{a priori}, we employ the same heuristic prescription used in Ref.~\cite{ho2020classical} to choose \(G\). Starting from an initial configuration sufficiently close to the sphaleron, standard gradient flow using Eq.~\eqref{eq:naiveGradientFlow} will minimise along the positive modes whilst continuing to flow the system along the negative mode. The point along the flow closest to the saddle point may be identified by considering the sum of the squares of the gradients at all lattice points. When this point is reached, the deviation from the true saddle point will be largely along the negative mode, and thus the modified flow \eqref{eq:chigusaFlow} is likely to converge along the saddle point. In practice, we found that flowing slightly further (\(\sim 1000\) iterations) along the negative mode from the point of closest approach gave better convergence. The normalised gradient at this point is used as \(G\) in our calculations:

\begin{equation}
G_\alpha=\frac{\partial_\alpha E_{\rm lat}}{\left(\sum_\beta \partial_\beta E_\mathrm{lat} \partial_\beta E_\mathrm{lat}\right)^{1/2}}.
\end{equation}

\bibliography{refs}

\end{document}